\documentstyle[aps,prb,epsf,multicol]{revtex}
\tighten
\begin{document}
\draft
\newcommand{\bn}{{\bf n}}
\newcommand{\bp}{{\bf p}}
\newcommand{\br}{{\bf r}}
\newcommand{\bq}{{\bf q}}
\newcommand{\bj}{{\bf j}}
\newcommand{\bE}{{\bf E}}
\newcommand{\eps}{\varepsilon}
\newcommand{\la}{\langle}
\newcommand{\ra}{\rangle}
\newcommand{\cK}{{\cal K}}
\newcommand{\cD}{{\cal D}}
\newcommand{\hp}{\hat p}
\newcommand{\hq}{\hat q}
\newcommand{\hx}{\hat x}
\newcommand{\hH}{{\hat H}_0}
\newcommand{\mybeginwide}{
     \end{multicols}\widetext
     \vspace*{-0.2truein}\noindent
     \hrulefill\hspace*{3.6truein}
}
\newcommand{\myendwide}{
     \hspace*{3.6truein}\noindent\hrulefill
     \begin{multicols}{2}\narrowtext\noindent
}
\title{\Large \bf{Effect of inelastic scattering on parametric pumping}
}

\author{M. Moskalets$^{1,2}$ and 
M. Buttiker$^1$
}
\address{
   $^1$D\'epartement de Physique Th\'eorique, Universit\'e de Gen\`eve,
   CH-1211 Gen\`eve 4, Switzerland\\
    $^2$Department of Metal and Semiconductor Physics,
        National Technic University "Kharkov Polytechnic Institute",
        Kharkov, Ukraine\\}
\date\today
\maketitle
\bigskip
\begin{abstract}
Pumping of charge in phase-coherent mesoscopic systems
due to the out-of-phase modulation
of two parameters
has recently found considerable interest.
We investigate the effect of inelastic processes on the
adiabatically pumped current through a two terminal mesoscopic sample.
We find that the loss of coherence
does not suppress the pumped charge but rather an additional
physical mechanism
for an incoherent pump effect comes into play.
In a fully phase incoherent system the pump effect is similar
to a rectification effect.

\bigskip\noindent
PACS numbers: 72.10.-d, 73.23.-b
\end{abstract}

\begin{multicols}{2}
\narrowtext

A recent experiment by Switkes et al. \cite{SMCG99}
generated
considerable interest in the adiabatic quantum pumping of charge.
In this experiment a dc voltage between two electron reservoirs
of an open quantum dot
was obtained by slow periodic but out-of-phase
variation of two independent parameters.
The dc potential difference is
a consequence of electron transfer between reservoirs.
The effect is discussed under conditions in which electron motion
is completely phase-coherent
and is therefore termed {\it quantum} pumping.
Here we are interested in the
effect of inelastic scattering. Interestingly, as we will show
inelastic scattering does not necessarily suppress the pumped charge but
can even enhance it.

There are several approaches to achieve charge pumping.
Perhaps the most direct way to generate charge transport
without applying an overall force to the system is to subject the system
to a traveling wave potential as envisaged by Thouless \cite{Thouless83}.
In that case there are conditions under which the pumped
charge is quantized \cite{Thouless83,Niu90}. The purely classical limit
is also intersting and has been used as a model for a
clocked Brownian computer \cite{RLMB}. Travelling potentials
can be generated for instance with the help of surface acoustic
waves \cite{Camb}.  Another possibility
is to exploit the Coulomb blockade effect
\cite{KJVHF91,PLUED92,OKKVH97,hazelzet} 
that leads to a quantization of a charge on
a quantum dot and is of metrological interest \cite{martinis}.
Still another possibility
\cite{SMCG99,SZB95,Brouwer98,ZSA99,SAA00,Simon,WWG00,LEWW00,AEGS00,Brouwer01,PB01,AEGS01},
called an adiabatic quantum electron pump, is based on an open dot
under conditions in which Coulomb blockade effects are not important.
In this case the pumped charge (either quantized or nonquantized)
depends on the interference of electron wave functions within the system.
It is this case which is of interest in this work. Inelastic
scattering destroies interference and we might naively expect
that the pump effect is suppressed as inelastic scattering increases.

An elegant formulation of the charge pumped
through a phase coherent conductor coupled to two
reservoirs with the same (electro-) chemical potentials
$\mu_l$=$\mu_r$=$\mu$ (see Fig.1) has been
given by Brouwer \cite{Brouwer98}. He derived a scattering matrix
expression for the pumped charge investigating the modulation
of the charge emissivity \cite{BTP94,Buttiker93}.
If some two parameters $X_1$, $X_2$ characterizing the system
are varied adiabatically and periodically
(with the frequency $\omega = 2\pi/\tau\to 0$)
the charge $Q$ transferred during one period
from the right reservoir to the left one is
\cite{Brouwer98}
\begin{equation}
   Q = e\int_0^\tau dt \left( \frac{dn_l}{dX_1} \frac{dX_1}{dt}
     + \frac{dn_l}{dX_2} \frac{dX_2}{dt} \right).
\label{Eq1}
\end{equation}
\noindent
Here $dn_l/dX$ is the emitance \cite{BTP94,Buttiker93}
into contact (lead) "l"
\begin{equation}
   \frac{dn_\alpha}{dX} = \frac{1}{2\pi} \sum_{\beta=l,r}
Im \left[\frac{\partial 
S_{\alpha\beta}}{\partial X} S^*_{\alpha\beta}\right],
\label{Eq2}
\end{equation}

\noindent where $S_{\alpha\beta}$ ($\alpha,\beta=l,r$)
are elements of a scattering matrix of a system under consideration
(note that because of a current conservation $Q$ may be expressed
through $-dn_r/dX$ in the same way).

The physical mechanism for adiabatic (parametric) pumping is as follows.
The infinitesimal change of system parameters $\delta X_i$
(for a time $\delta t$) leads to a redistribution of the charge
within (and around) the system. The redistribution of the charge is
a consequence of the variation of the electron
density of states and produces electron flows through the 
system \cite{Brouwer98}
$I_i(X_1,X_2)$=$\delta Q_i/\delta t$, where
$\delta Q_i(X_1,X_2)$=$e dn/dX_i\delta X_i$, ($i=1,2$). 
These currents (and the pumping effect)
are thus a consequence of the changing electrostatic landscape.

The effect of inelastic interactions may be
understood in the following way. To be definite let us
consider a quantum dot
coupled to reservoirs via two point contacts.
Without inelastic interactions
electrons pass coherently from one reservoir to another and
the pumped charge shows a resonant like behavior
\cite{WWG00,LEWW00} that is a consequence of interference 
between the two point contacts. The inelastic interactions within 
a quantum
dot destroy this quantum interference and allow us to consider
the dot's interior as
a third reservoir \cite{Buttiker88}
with some potential $\mu_f (t)$ which is determined such that
the net current
into this additional reservoir is zero at every instant.

\begin{figure}
\vspace{3mm}
\centerline{
\epsfxsize8cm
\epsffile{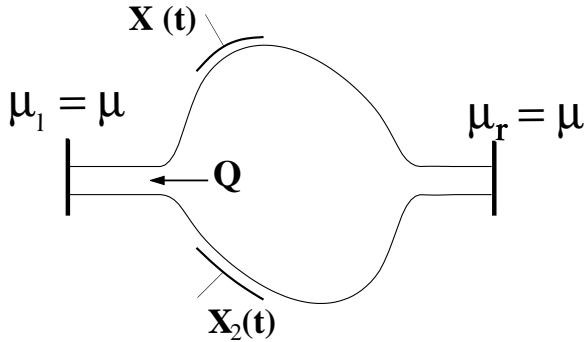}
}
\nopagebreak
\caption{
A mesoscopic system coupled to two reservoirs
with the same potentials $\mu_l=\mu_r=\mu$.
When the parameters $X_1$ and $X_2$ are changed periodically but out 
phase a charge $Q$ can be transferred from one reservoir to another one.
}
\label{FIG.1a}
\end{figure}
\noindent 
For a typical geometry the currents pumped through the two contacts
are unequal and some charge is pumped
to the third reservoir generating a time-dependent
potential $\mu_f (t)$.
Due to this potential additional currents flow into (or from)
the reservoirs. The existence of these additional currents
is at the origin of an inelastically driven electron pump.
Thus in the presence of inelastic interactions there is a
novel rectification mechanism which as we will show can give
rise to a dc current. We emphasize already here, that this effect differs
from the rectification of displacement currents recently discussed
by Brouwer \cite{Brouwer01} and Polianski and Brouwer \cite{PB01}
which is closely related to the actual set-up of the experiment
of Switkes et al. \cite{SMCG99}.

To illustrate the inelastic rectification mechanism giving raise
to a pumping effect
we consider
a generic case: a point scatterer with an oscillating strength coupled to
two reservoirs with a periodic in time chemical potential difference
$\delta\mu (t) $ = $\mu_l$ - $\mu_r$.
The time-modulation of the point scatterer
leads to a conductance
\begin{equation}
   G(t) = G_0 + \delta G \sin(\omega t),
\label{Eq3}
\end{equation}
\noindent
and the chemical potential difference is
\begin{equation}
   \delta\mu(t) =  \delta\mu\sin(\omega t + \varphi).
\label{Eq4}
\end{equation}
with $\varphi$ an arbitrary phase.

\begin{figure}
  \vspace{3mm}
  \centerline{
   \epsfxsize8cm
   \epsffile{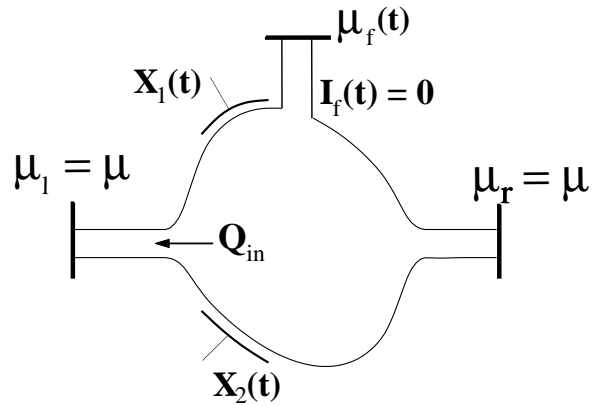}
  }
  \nopagebreak
  \caption{
To introduce inelastic interactions we couple the system (Fig.1)
to the third (fictitious) reservoir with a time-dependent chemical
potential $\mu_f(t)$. The net current into this reservoir
is zero at any time $I_f(t)=0$.
  }
\label{FIG.1b}
\end{figure}
\noindent
The charge $Q$ transferred through the sample in one period
$\tau=2\pi/\omega$ is $Q$=$\int_0^\tau dt G(t)\delta\mu(t)$. Using
Eqs.(\ref{Eq3}),(\ref{Eq4}) we find a dc current $I_{dc}$=$Q/\tau$
pumped through the sample given by
\begin{equation}
   I_{dc} = \frac{1}{2}\delta G\delta\mu\cos(\varphi).
\label{Eq5}
\end{equation}

\noindent
Note, that this pump driven by rectification has similar features
as the quantum coherent adiabatic pump.
Namely, this effect exists if at least two system parameters are varied
in time. In the above example these parameters are the point scatterer
(the conductance) and the inelastically generated
potential difference.
In addition, the pumped current $I_{dc}$ given by Eq.(\ref{Eq5}) 
is proportional
to the frequency $\omega$. This follows from the fact
(see the discussion after Eq.(\ref{Eq7}) that in the adiabatic pump
considered here the potential difference
(produced by inelastic interactions)
$\delta\mu\sim I_i$ $\sim dX_i/dt\sim \omega$
is also proportional to the pump frequency $\omega$.
So, we have $I_{dc}\sim\omega$ as it must be for a pump effect.

Below we present a more formal derivation which allows us to
consider the effect of inelastic interactions on the pump effect
within a scattering approach.
To introduce inelastic interactions into the system under consideration we
will use the model of Ref. \onlinecite{Buttiker88}
in which a third, fictitious voltage probe is coupled to 
the conductor (see Fig.2). We can only give a few recent
references to indicate the breadth of questions addressed 
with this approach \cite{JB96}. 
The net current $I_f(t)$
between the system and this additional 
reservoir is taken to be zero at any time
(i.e., the charge leaving the system
through the third lead for the time $\delta t$ is zero,
$\delta Q_f(t)=I_f(t)\delta t=0$).
Note, that this assumption leads to a time dependent electrochemical
potential difference between the fictitious reservoir and the system
(and, respectively, between fictitious and real reservoirs).
Some carriers are scattered into the fictitious
reservoir and the resulting current is compensated by carriers
injected from the reservoir into the system.
Carriers entering and leaving a reservoir have
a phase and energy which is unrelated.
Physically the third reservoir
describes thus carriers which undergo an
inelastic scattering process within the mesoscopic
sample and as a consequence loose their phase coherence.

The presence of the third lead modifies the charge $Q_\alpha$
leaving the sample through the leads $\alpha=l,r$.
After straightforward calculations taking into account that $\delta Q_f(t) =0$
we get a pumped charge $Q_{in}$ due to
inelastic
interactions of the following form
\begin{eqnarray}
   Q_{in}=Q_{in,l}=-Q_{in,r};~~~~~~~~~~~~~~~~~~~~~~~~~~~~~~ \nonumber \\
   Q_{in,\alpha} = e\int_0^\tau dt \left(
F(\alpha) + K_{in,\alpha} F(f)\right),~~~
                      \alpha=l,r,
\label{Eq6}  \\
   F(\gamma)=\frac{dn_{in,\gamma}}{dX_1}\frac{dX_1}{dt} +
            \frac{dn_{in,\gamma}}{dX_2}\frac{dX_2}{dt},~~~
               \gamma=l,r,f. \nonumber
\end{eqnarray}

\noindent Here
the emissivities $dn_{in,\gamma}/dX$ ($\gamma=l,r,f$) are
\begin{equation}
   \frac{dn_{in,\gamma}}{dX} = \frac{1}{2\pi} \sum_{\beta=l,r,f}
Im \left[\frac{\partial S_{\gamma\beta}}{\partial X} 
S^*_{\gamma\beta} \right].
\label{Eq7}
\end{equation}
\noindent
In Eq. (\ref{Eq7}) the index of summation now also runs over the
fictitious lead.
The coefficients $K_{in,\alpha}=\frac{T_{f\alpha}}{T_{fl}+T_{fr}}$
(where $T_{f\alpha}=|S_{f\alpha}|^2$ is the probability for
electrons entering the sample through the lead $\alpha=l,r$ to be
inelastically scattered within the sample)
describe a redistribution of outgoing particles
undergoing inelastic processes within the system between
the left and the right leads.

In the above expression the second term contains a new
(compared with a purely coherent case) physical mechanism for
an adiabatic electron pump. This mechanism is a rectification
of incoherent currents flowing into the system
from the left and the right reservoirs (or vice versa).
When the system parameters $X_1$ and $X_2$
are varied the current flowing into the third (fictitious) reservoir
(i.e., an incoherent current) is $I_{in}(t)=eF(f)$.
To avoid a charge accumulation within the system (within the third reservoir)
this current induces an additional potential
$\delta\mu(t)$=$\mu_f(t)-\mu$=
$I_{in}(t)/G$ (where $G=e^2/h(T_{fl}+T_{fr})$) in the system
interior in order to maintain the full current
$I_f$ into this reservoir at zero.
The presence of a potential $\delta\mu(t)$ changes the currents flowing
into the real reservoirs by amount
$\Delta I_{\alpha}(t)=e^2/h T_{f\alpha}\delta\mu=I_{in}(t)K_{in\alpha}(t)$.
Integrating $\Delta I_{\alpha}$ over a time period $\tau$ we get
an additional charge pumped through the system. This charge has
a rectified part that survives even if the coherent pump effect is
fully destroyed by inelastic interactions.

To provide an example we now consider the
simplest system which can pump a charge.
As was shown by Brouwer \cite{Brouwer98} parametric pumping may
occur if at least two (independent) system parameters are varied. From
this it immediately follows that a point scatterer
(a delta function potential)
with a potential $U_1(x,t)=V(t)\delta(x)$
does not show a pump effect, because in this case
the scattering is characterized by one parameter only - the strength $V$. 
Thus, only a partially extended system can
pump a charge.
Thus we consider
two point scatterers located at some distance from each other
$U(x,t)$ = $V_l(t)\delta(x+a)$ + $V_r(t)\delta(x-a)$.
The pump effect in such a coherent system was considered in
\cite{WWG00,LEWW00}.
Inelastic interactions destroy coherent tunneling and
lead to a sequential (incoherent) tunneling through two barriers
\cite{Buttiker88}.

To introduce inelastic interactions we couple the system to
the third reservoir using a wave-splitter located at $x=0$ described by
a scattering matrix $S_e$ \cite{Buttiker88}
\begin{equation}
S_e=\left(
  \begin{array}{cccc}
    0 & \sqrt{1-\epsilon} & \sqrt{\epsilon} & 0   \\
    \sqrt{1-\epsilon} & 0 & 0 & \sqrt{\epsilon}    \\
    \sqrt{\epsilon}   & 0 & 0 & -\sqrt{1-\epsilon}  \\
    0 & \sqrt{\epsilon} & -\sqrt{1-\epsilon} & 0   \\
  \end{array}
\right)
\label{Eq8}
\end{equation}

\noindent
Here the coupling parameter $\epsilon$ characterizes the strength
of inelastic interactions. At $\epsilon=1$ all electrons are inelastically
scattered within the system.
Whereas at $\epsilon=0$ electrons move
coherently through the system (from channel 1 to channel 2) and a fictitious
reservoir (channels 3 and 4) fully decoupled from the system under
consideration.

First, we consider the limit of strong inelastic interactions $\epsilon=1$.
In this case we have no coherent tunneling \cite{Buttiker88}
and the system may be considered as two classical resistors
in series
(point potential barriers with heights $V_\alpha(t)$, $\alpha=l,r$)
with the scattering matrices  $S_{11}^{(\alpha)}$=$S_{22}^{(\alpha)}$=
$-i\xi_\alpha/(1+i\xi_\alpha)$; $S_{12}^{(\alpha)}$=$S_{21}^{(\alpha)}$=
$1/(1+i\xi_\alpha)$, where $\xi_\alpha$=$V_\alpha/(\hbar v)$ and
$v$ is an electron velocity.
To avoid a misunderstanding note, that at $\epsilon=1$
all electrons entering the system through the channel 1 (2) go into
channel 3 (4) (and vice versa). Thus, the system containing four leads
(two real channels 1 and 2 and two fictitious channels 3 and 4)
is divided into two subsystems (the scatterers $V_l$ and $V_r$)
with two leads only (channels

\begin{figure}
  \vspace{3mm}
  \centerline{
   \epsfxsize8cm
   \epsffile{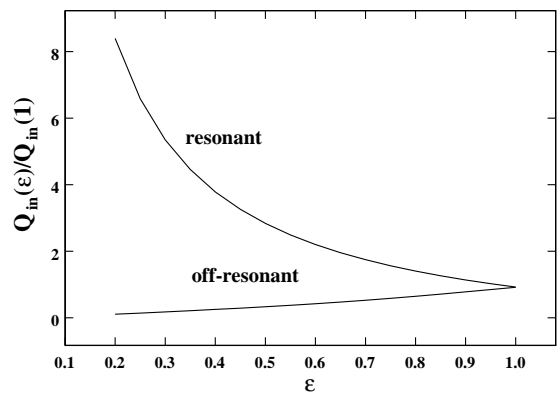}
  }
  \nopagebreak
  \caption{
The dependence on the strength $\epsilon$
of inelastic interactions of the charge $Q_{in}$ pumped through
a double-barrier potential
$U(x)=V_l(t)\delta(x+d)+V_r(t)\delta(x-d)$
at small variations of the potential strengths 
$V_\alpha(t)=V_0+\delta V_\alpha(t)$, $\delta V_\alpha\ll V_0$,
$\alpha=l,r$.
The parameters are: $V_0/(\hbar v)=5$; $k_F d=3.04$ modulo $\pi$ (resonant);
$k_F d=\pi/4$ modulo $\pi$ (off-resonant).
Here $v, k_F$ are the electron velocity and Fermi wave vector.
  }
\label{FIG.2}
\end{figure}
\noindent 
(1,3) and (2,4), respectively).
Though each of the subsystems has only one parameter ($V_\alpha(t)$)
which is varied (and, consequently, they do not show a coherent pump effect)
the full system can pump current due to a rectification mechanism.
Substituting $K_{in,l}$=$G_l/(G_l+G_r)$ and the scattering
matrix $S^{\alpha}$ into Eqs.(\ref{Eq6}) and (\ref{Eq7}) we get a pumped charge
\begin{equation}
   Q_{in} =-\frac{1}{ev} \int_0^\tau dt
       \left( G_l^{-1}(t)+G_r^{-1}(t)\right)^{-1}
       \left\{  \frac{\partial V_l}{\partial t}
           - \frac{\partial V_r}{\partial t}
       \right\}.
\label{Eq9}
\end{equation}

\noindent
Here $G_\alpha$=$(e^2/h) |S_{12}^{\alpha}|^2$.

In the limit of a small modulation of the height of the potential barriers
$V_{\alpha}(t)$ = $V_{0\alpha}$ +
$\delta V_\alpha\sin(\omega t+\varphi_\alpha)$,
$\delta V_\alpha\ll V_{0\alpha}$ ($\alpha=l,r$)
for equal opaque barriers
$V_{0l}=V_{0r}=V_0$ and $V_0\gg \hbar v$
we get a charge
pumped for one period $\tau$
\begin{equation}
   Q_{in} = e\sqrt{T_0}\frac{\delta G_l\delta G_r}{8G_0^2}
   \sin(\varphi_r - \varphi_l),
\label{Eq10}
\end{equation}

\noindent
where $G_0=e^2/h T_0$; $T_0=(\hbar v/V_0)^2$ and
$\delta G_\alpha$=$-2G_0\delta V_\alpha/V_0$.

Let us compare $Q_{in}$ with the pumped charge \cite{LEWW00}
in the fully coherent limit.
For a small variation $\delta V_\alpha\ll V_0$
the off-resonant pumped charge is
$Q_{off}\sim eT_0\sqrt{T_0}\delta G_l\delta G_r/G_0^2$.
Thus away from a transmission resonance the inelastic interactions
increase the pumped charge by a factor of $\sim 1/T_0$.

Near a resonance it is more convenient to consider a large pumped
cycle which encloses a resonance line \cite{LEWW00}.
In this case the coherent pumped charge is quantized.
The inelastic interactions destroy the Fabry-Perot interference effect and
destroy the quantization of the pumped charge. In particular,
for a limit cycle which encloses
(counterclockwise) a region $0<V_l,V_r<\infty$
the equation (\ref{Eq9}) gives $Q_{in}$=$e/2^{3/2}$.

These results are illustrated in Fig.3 which shows 
the effect of inelastic scattering away from resonance 
and on resonance as a function of the coupling strength $\epsilon$
to the fictitious voltage lead.  
Whereas near a resonance inelastic scattering
suppresses the pumped charge (we assume
that the potential variations do not lead us away from 
the resonant region, i. e. a small amplitude pump cycle) 
the off-resonant pump current is increased 
by inelastic scattering. 
Note, that the inelastic interactions affect
the conductance of a double barrier structure
in an analogous way \cite{Buttiker88}.

In conclusion, we have shown that inelastic interactions
within a mesoscopic system introduce
an additional mechanism which contributes to adiabatic electron pumping.

M.M. appreciates the warm hospitality of
the Department of Theoretical Physics of the University of Geneva,
where part of this work was done. This work was supported by the Swiss
National Science Foundation.

\end{multicols}
\end{document}